\documentclass[twocolumn,floats,prb,aps]{revtex4}

\usepackage{graphicx}
\usepackage{dcolumn} 
\usepackage{amsmath}
\usepackage{exscale}
\usepackage{bm}      
\usepackage{color}
\usepackage{epsfig,psfrag,subfigure,subeqnarray}
\usepackage{bbm}
\usepackage{amsbsy}
\usepackage{amssymb}
\usepackage{enumitem}
\usepackage{mathrsfs}
\usepackage{amsfonts}

\def\lan{\langle}
\def\ran{\rangle}

\def\vk{{\bf k}}

\def\vr{{\bf r}}
\def\vL{{\bf L}}
\def\vJ{{\bf J}}
\def\vS{{\bf S}}

\def\vQ{{\bf Q}}

\def\vp{{\bf p}}
\def\vR{{\bf R}}

\def\v0{{\bf 0}}
\newcommand{\bd}{\begin{equation}}
\newcommand{\ed}{\end{equation}}
\newcommand{\be}{\begin{equation}}
\newcommand{\ee}{\end{equation}}
\newcommand{\bt}{\begin{split}}
\newcommand{\et}{\end{split}}

\newcommand{\bn}{\begin{align}}
\newcommand{\en}{\end{align}}

\newcommand{\bea}{\begin{eqnarray}}
\newcommand{\eea}{\end{eqnarray}}
\newcommand{\ba}{\begin{array}}
\newcommand{\ea}{\end{array}}
\newcommand{\nn}{\nonumber}

\DeclareMathAlphabet\mathbfcal{OMS}{cmsy}{b}{n}

\begin{document}

\title{From spherical to periodic symmetry:\\
the analog of  orbital angular momentum  for semiconductor crystals}

\author{Monique Combescot$^1$, and  Shiue-Yuan Shiau$^2$}\
\affiliation{(1) Institut des NanoSciences de Paris, Sorbonne Universit\'e, CNRS, 4 place Jussieu, 75005 Paris}
\affiliation{(2) Physics Division, National Center for Theoretical Sciences, Taipei, 10617, Taiwan}

\begin{abstract}
The angular momentum formalism provides a powerful way to classify atomic states. Yet, requiring a spherical symmetry  from the very first line, this formalism cannot be used for periodic systems, even though cubic semiconductor states are commonly classified according to atomic notations. Although never noted, it is possible to define the analog of the orbital angular momentum, by only using the potential felt by the electrons. The spin-orbit interaction for crystals then takes the $\mathbfcal{\hat{L}}\cdot \hat{\vS}$ form, with $\mathbfcal{\hat{L}}$ reducing to $\hat{\vL}=\vr\times\hat{\vp}$ for spherical symmetry. This provides the long-missed support for using the eigenvalues of $\mathbfcal{\hat{L}}$ and $\mathbfcal{\hat{J}}=\mathbfcal{\hat{L}}+\hat{\vS}$, as quantum indices to label cubic semiconductor states. Importantly, these quantum indices also control the phase factor that relates valence electron to hole operators, in the same way as particle to antiparticle, in spite of the fact that the hole is definitely not the valence-electron antiparticle. Being associated with a broader definition, the ($\mathbfcal{\hat{L}},\mathbfcal{\hat{J}}$) analogs of the $(\hat{\vL},\hat{\vJ})$ angular momenta, must be distinguished by names: we suggest ``spatial momentum" for $\mathbfcal{\hat{L}}$ that acts in the real space, and ``hybrid momentum" for $\mathbfcal{\hat{J}}$ that also acts on spin, the potential symmetry being specified as ``cubic spatial momentum". This would cast $\hat{\vJ}$ as a ``spherical hybrid momentum", a bit awkward for the concept is novel.

\end{abstract}
\date{\today}
\maketitle

It is well established\cite{Landau01,Sakurai,Cohenphotonatom,Pethick} that the state classification for electrons in a spherical potential, \textit{i.e.,} a potential that only depends on the modulus of the electron coordinate $\vr$, is driven by the orbital angular momentum operator 
\be
\hat{\vL}=\vr\times\hat{\vp}\label{1}
\ee
where $\hat{\vp}=-i\hbar \nabla$ is the electron momentum operator.
This operator controls the interplay between the spin and the spatial degree of freedom of atomic electrons, via the spin-orbit interaction\cite{LS} that for a spherical potential, takes the simple $\hat{\vL}\cdot \hat{\vS}$ form. This scalar product is handled by introducing the angular momentum operator $\hat{\vJ}=\hat{\vL}+\hat{\vS}$. Since $\hat{\vL}\cdot \hat{\vS}$ is equal to $(\hat{\vJ}^2-\hat{\vL}^2-\hat{\vS}^2)/2$, while $\hat{\vL}^2$ and $\hat{\vS}^2$ commute with the atomic Hamiltonian, the derivation of the spin-orbit eigenstates just amounts to deriving the $\hat{\vJ}^2$ eigenstates, which is easy to do with the help of the angular momentum formalism. Although mathematically smart, the concept of angular momentum is physically odd because $\hat{\vJ}$ is the sum of two operators that act in different subspaces. Yet, the angular momentum formalism provides a very elegant way to derive the degeneracy of the various atomic states.

While $\hat{\vL}$ commutes with the Hamiltonian for atomic electrons in a spherical potential, this is not so for semiconductor electrons in a periodic potential\cite{VAKu}; so, the orbital angular momentum $\vL$ would not be a conserved quantity. Moreover, this quantity would be infinite in the large sample limit: indeed, semiconductor electrons have a wave function that extends over the whole sample\cite{Kittelbook,Merminbook}. The mean value of the $\vr$ coordinate for such ``itinerant'' electron goes to infinity, whereas it stays finite for a bound electron orbiting around a nucleus, whatever the sample size.

  Still, valence electron states in a Zinc-Blende-like cubic crystal are commonly labeled along the atomic notations, which is fully questionable from the very first line because these notations are related  to problems having a spherical symmetry. The valence electrons of such crystals have a threefold spatial degeneracy\cite{Cardona}, labeled along the crystal axes $(x,y,z)$. Like atomic electrons, they suffer a spin-orbit interaction that split this spatial degeneracy\cite{Dresselhaus1955,KL,Luttinger,Warren} in a way that should be similar with how the threefold $\ell=1$ atomic level splits, because the $(x,y,z)$ cyclic permutations fundamentally corresponds to a spherical symmetry ``restricted to" a $\pi /2$ rotation. Consequently, even if semiconductor electrons do not have an orbital angular momentum due to the lack of spherical symmetry, they should have a vector operator that plays a similar role. 
 
 \textbf{The goal of the present work} is to identify the vector operator for periodic potentials that is conceptually similar to the orbital angular momentum operator for  a spherical potential, and to possibly classify the energy levels of semiconductor electrons without borrowing atomic notations that rely on a concept invalid for a periodic system. To do it, we go back to the foundation of the spin-orbit interaction, in order to follow the spirit of the procedure that leads to the classification of atomic states.

 The general form of the spin-orbit interaction for \textit{whatever the system symmetry}, reads\cite{Thomas}
\be\label{2}
\hat{H}_{so}=\lambda_{so}\Big(\nabla\mathcal{V}(\vr)\times\hat{\vp}\Big)\cdot \hat{\vS}
\ee  
where $\lambda_{so}$ is a constant and $\mathcal{V}(\vr)$ is the internal electrostatic potential felt by the electrons. This interaction splits into an operator $(\nabla\mathcal{V}(\vr)\times\hat{\vp})$ that only acts in the real space, and $\hat{\vS}$ that only acts in the spin space. 

For a spherical potential, as in the case of atomic electrons, the operator that acts on space reduces to
\bea\label{3}
\hat{\bf\Lambda}^{(at)}&=&\nabla \mathcal{V}^{(at)}(\vr)\times \hat{\vp}=\left(\frac{d}{dr}\mathcal{V}^{(at)}(r)\,\,\,\frac{\vr}{r}\right) \times \hat{\vp}\\
&=&\lambda_{at}(r)\,\hat{\vL}\nn
\eea
from which the usual $\hat{\vL}\cdot \hat{\vS}$ form for the spin-orbit interaction follows  readily.

When the potential is periodic, as in the case of  crystals, we can, by analogy, introduce the vector operator $\hat{\mathbfcal{L}}$ defined through\cite{MonicPRB2019}
\be\label{4}
\hat{\bf\Lambda}=\nabla \mathcal{V}(\vr)\times \hat{\vp}\equiv\lambda(\textbf{r})\,\hat{\mathbfcal{L}}
\ee
 for the spin-orbit interaction to take a similar $\hat{\mathbfcal{L}}\cdot\hat{\vS}$ form. The $\lambda(\textbf{r})$ scalar follows from enforcing $\hat{\mathbfcal{L}}$ to have the same dimensionality as $\hat{\vL}$, that is, eigenvalues that scale as $\hbar$.

The vector operator $\hat{\mathbfcal{L}}$ reduces to $\hat{\vL}=\vr\times\hat{\vp}$ when the $\mathcal{V}(\vr)$ potential has a spherical symmetry; but it is fundamentally different in the case of an itinerant electron in a periodic crystal; such electron does not \textit{orbit} around a particular  nucleus; so, its position can not be described by a distance $r$ plus two angles $(\theta, \varphi)$. Still, we will show that, in the case of a \textit{cubic crystal}, the matrix representation of the corresponding $\hat{\mathbfcal{L}}^{(cb)}$ operator in the threefold basis $(x,y,z)$ taken
 along the crystal axes, is identical to the one of the orbital angular momentum $\hat{\vL}$ in the threefold basis 
 of the $\ell=1$ atomic level, provided that the $(X,Y,Z)$ atomic axes, that are arbitrary for spherical symmetry, are chosen along the cubic crystal axes. As a direct consequence, the spin-orbit interaction  $\hat{\mathbfcal{L}}^{(cb)}\cdot\hat{\vS}$ must split the threefold spatial level of electrons in a cubic crystal, in exactly the same way as $\hat{\vL}\cdot\hat{\vS}$ does for the $\ell=1$ atomic level: we just have to introduce the vector operator $\hat{\mathbfcal{J}}^{(cb)}=\hat{\mathbfcal{L}}^{(cb)}+\hat{\vS}$; its eigenstates will give the spin-orbit eigenstates for cubic semiconductors, just as the angular momentum $\hat{\vJ}$ resolves the spin-orbit eigenstates of atomic electrons.

Yet, the difference that exists between $\hat{\mathbfcal{L}}$ defined in Eq.~(\ref{4}) for whatever potential symmetry, and $\hat{\vL}$ defined in Eq.~(\ref{1}), calls for different names. The vector $\hat{\vL}$ has been named ``orbital angular momentum'' because the atomic electron is best described by two \textit{angles} and a distance with respect to the nucleus around which it \textit{orbits}. It is clear that neither \textit{orbital} nor \textit{angular} can be used to qualify $\hat{\mathbfcal{L}}$ because these words are too closely related to a spherical symmetry. We are left with what the operator $\hat{\mathbfcal{L}}$ is: it acts in the real space. This is why we suggest to call it ``spatial momentum''. By adding the spin to get $\hat{\mathbfcal{J}} =\hat{\mathbfcal{L}}+\hat{\vS} $, we construct a hybrid operator that acts in two subspaces. We suggest to call it ``hybrid momentum". Of course, these two names encompass $\hat{\vL}$ and $\hat{\vJ}$, just as when a broader concept is introduced in physics. Still, the precise matrix forms of these operators depend on the symmetry of the problem. This is why it is necessary to add this symmetry to their name: $\hat{\mathbfcal{L}}^{(cb)}$ for cubic crystals should be named ``cubic spatial momentum", while the ``orbital angular momentum" $\hat{\vJ}$ is a ``spherical spatial momentum'' within this broader understanding. Note that since $\hat{\textbf{S}}$ has a spherical symmetry, the operators $\hat{\mathbfcal{L}}$ and $\hat{\mathbfcal{J}}$ have the same symmetry.

The paper is organized as follows.

In Section \ref{sec2}, we reconsider the spin-orbit interaction in the case of atoms, to settle the procedure. 

In Section \ref{sec3}, we consider the $\hat{\mathbfcal{L}}$ operator defined in Eq.~(\ref{4}), and we calculate its matrix representation in the threefold valence-electron subspace when the periodic potential has a cubic symmetry. We show, by using the spatial basis $|v,\ell_z\ran$ with $\ell_z=(\pm1,0)$ instead of $|v,\mu\ran$ with $\mu=(x,y,z)$, that the $\hat{\mathbfcal{L}}^{(cb)}$ operator has exactly the same matrix elements as the orbital angular momentum $\hat{\vL}$ in the $\ell=1$ subspace.  As a direct consequence, the matrix elements of the ``cubic hybrid momentum'' $\hat{\mathbfcal{J}}^{(cb)}=\hat{\mathbfcal{L}}^{(cb)}+\hat{\vS}$ in the sixfold spin-orbit eigenstate subspace are just the same as the ones of the angular momentum $\hat{\vJ}$ in the six $j=(3/2,1/2)$ atomic states. As a result, the operators $\hat{\mathbfcal{J}}^{(cb)}$ and $\hat{\vJ}$ have the same eigenvalues and their eigenstates have the same forms. This brings the long-missed support for using atomic notations for cubic semiconductor.

In Section \ref{sec4}, we  show that although the valence hole is for sure not a na\"{i}ve antiparticle, the relation between the destruction operator of a particle  and the creation  operator of its antiparticle\cite{Landau01}, established in relativistic quantum theory\cite{Berestetskii}, stays valid for valence electron and hole in a \textit{cubic semiconductor}, on condition that we label the valence electrons along quantum indices $\mathcal{L}$ or $\mathcal{J}$ that correspond to the $\hat{\mathbfcal{L}}^{(cb)}$ or $\hat{\mathbfcal{J}}^{(cb)}$ eigenvalues. This provides a secure way for turning from valence electron operator to hole operator when writing the semiconductor physics in terms of electrons and holes, as it should for problems dealing with semiconductor excitations.

In Section \ref{sec5}, we discuss the possibility to extend the $\hat{\mathbfcal{L}}$ concept defined in Eq.~(\ref{4}), that we here use for electrons in a cubic semiconductor, to crystals having a different symmetry. This should provide a smart way to derive the spin-orbit eigenstates for semiconductor crystals, without relying on the physically obscure group theory formalism. 

We then conclude.

\section{Spin-orbit interaction for atoms\label{sec2}}

\subsection{Angular momentum formalism}
We  consider an electron in the $|\ell,\ell_Z\ran$ eigenstate of the operators $\hat{\vL}^2$ and $\hat{L}_Z$, with $\ell=1$ and $\ell_Z=(\pm1,0)$ for a threefold orbital level, the quantization axis $\bf Z$  being chosen at will due to spherical symmetry.  The eigenstates of the operators $\hat{\vJ}^2$ and $\hat{J}_Z$, with $\hat{\vJ}=\hat{\vL}+\hat{\vS}$, are $|j,j_Z\ran$ with $j=\ell\pm1/2$ and $j_Z=(j,j-1,\cdots,-j)$. From $\hat{\vJ}^2|j,j_Z\ran=\hbar^2j(j+1)|j,j_Z\ran$ and similar relations for $\hat{\vL}^2$ and $\hat{\vS}^2$, we readily find that $\hat{\vL}\cdot\hat{\vS}=(\hat{\vJ}^2-\hat{\vL}^2-\hat{\vS}^2)/2$ acting on the state $\left|j=3/2,j_Z\right\ran$ made of $\ell=1$ orbital states and $s=1/2$ spin states, reads as
\bea
\hat{\vL}\cdot\hat{\vS}\left|j=\frac{3}{2},j_Z\right\ran&=&\frac{\hbar^2}{2}\left(\frac{3}{2}\cdot \frac{5}{2}-1\cdot2-\frac{1}{2}\cdot\frac{3}{2}\right)\left|\frac{3}{2},j_Z\right\ran\nn\\
&=& \frac{\hbar^2}{2}\left|\frac{3}{2},j_Z\right\ran\label{5}
\eea
In the same way, $\hat{\vL}\cdot\hat{\vS}\left|j=1/2,j_Z\right\ran=-\hbar^2 \left|1/2,j_Z\right\ran$. Since the spin-orbit interaction for a spherical potential is proportional to $\hat{\vL}\cdot\hat{\vS}$, the above results show that this interaction splits the $(3\times2)$-fold atomic level  $(\ell_Z,s_Z)$ into a fourfold level  $(j=3/2,j_Z)$ and a twofold level  $(j=1/2,j_Z)$.

\subsection{Matrix representations}
An easy way to relate atomic operator to cubic semiconductor operator is through their matrix representations. To possibly do it, let us recall the matrix forms of $\hat{\vL}$ and $\hat{\vJ}$.

\noindent \textbf{(a) Orbital angular momentum} $\hat{\vL}$

The basis for the threefold atomic level $\ell=1$ corresponds to $|\ell_Z\ran$  with $\ell_Z=(\pm1,0)$. The angular momentum formalism tells that the operators  $\hat{L}_Z$ and $\hat{L}_\pm=\hat{L}_X\pm i\hat{L}_Y$  are such that $\hat{L}_Z|\ell_Z\ran=\hbar\ell_Z |\ell_Z\ran$ and $\hat{L}_\pm|\ell_Z\ran=\hbar \sqrt{1\cdot2-\ell_Z(\ell_Z\pm 1)}\,\,|\ell_Z\pm 1\ran$. This gives the matrices representing these operators in the $\ell_Z=(1_Z,0_Z,-1_Z)$ basis, as
\be\label{6}
\hat{L}_Z=\hbar\left(\begin{matrix}
1 & 0 & 0 \\
0& 0 & 0  \\
0 & 0 & -1
\end{matrix}\right)_{1;Z} \quad\hat{L}_+=\hat{L}^\dag_-=\hbar \sqrt{2}\left(\begin{matrix}
0 & 1 &0 \\
0& 0 &1\\
0& 0 & 0
\end{matrix}\right)_{1;Z} 
\ee
from which we get
\be\label{7}
\hat{L}_X=\frac{\hbar}{\sqrt{2}}\left(\begin{matrix}
0 & 1 &0 \\
1& 0 &1\\
0& 1 & 0
\end{matrix}\right)_{1;Z}\quad \hat{L}_Y=\frac{\hbar}{\sqrt{2}}\left(\begin{matrix}
0 & -i & 0\\
i& 0 & -i \\
0 & i & 0
\end{matrix}\right)_{1;Z}
\ee


We can check that $\left[ \hat{L}_X, \hat{L}_Y\right]_-=i\hbar \hat{L}_Z$ and $\hat{\vL}^2= \hat{L}_X^2+ \hat{L}_Y^2+ \hat{L}_Z^2=(1\cdot 2)\hbar^2\,\hat{\textrm{I}}^{(3)}$, where $\hat{\textrm{I}}^{(3)}$ is the $3\times3$ identity matrix.

\noindent \textbf{(b) Angular momentum} $\hat{\vJ}$

The angular momentum formalism gives the $\hat{\vJ}$ matrices in the $j=3/2$ eigenstate basis $|j_Z\ran$ with $j_Z=(\pm3/2,\pm1/2)$, through $\hat{J}_Z|j_Z\ran=\hbar j_Z|j_Z\ran$ and $\hat{J}_\pm|j_Z\ran=\hbar \sqrt{3/2\cdot5/2-j_Z(j_Z\pm 1)}\,\,|j_Z\pm 1\ran$. We then find in the $j_Z=(3/2_Z,1/2_Z,-1/2_Z,-3/2_Z)$ basis
\be
\hat{J}_Z{=}\hbar\left(\begin{matrix}
\frac{3}{2} & 0 & 0 & 0 \\
0& \frac{1}{2} & 0 & 0 \\
0 & 0 & {-}\frac{1}{2} & 0 \\ 
0 & 0 & 0 & {-}\frac{3}{2}
\end{matrix}\right)_{\frac{3}{2};Z}\,\, \hat{J}_+{=}\hat{J}_-^\dag{=}\hbar\left(\begin{matrix}
0& \sqrt{3} & 0 & 0 \\
0&0 & 2 & 0 \\
0 & 0 & 0 & \sqrt{3} \\ 
0 & 0 & 0 & 0
\end{matrix}\right)_{\frac{3}{2};Z}\label{10}
\ee
from which we get, for $\hat{J}_\pm=\hat{J}_X\pm i \hat{J}_Y$,
\bea
\hat{J}_X=\hbar\left(\begin{matrix}
0 & \frac{\sqrt{3}}{2} &0 & 0 \\
\frac{\sqrt{3}}{2}& 0 &1 & 0\\
0& 1 & 0 & \frac{\sqrt{3}}{2} \\
0 & 0 & \frac{\sqrt{3}}{2} & 0
\end{matrix}\right)_{\frac{3}{2};Z}\nn\\ \hat{J}_Y=\hbar\left(\begin{matrix}
0 & -i\frac{\sqrt{3}}{2} &0 & 0 \\
i\frac{\sqrt{3}}{2} & 0 &-i & 0\\
0& i & 0 & -i\frac{\sqrt{3}}{2} \\
0 & 0 & i\frac{\sqrt{3}}{2} & 0
\end{matrix}\right)_{\frac{3}{2};Z}\label{11}
\eea

As for $\hat{\vL}$, the $\hat{\vJ}$ operator fulfills $\left[ \hat{J}_X, \hat{J}_Y\right]_-=i\hbar \hat{J}_Z$ and $\hat{\vJ}^2= \hat{J}_X^2+ \hat{J}_Y^2+ \hat{J}_Z^2=\left(\frac{3}{2}\cdot \frac{5}{2}\right)\hbar^2\,\hat{\textrm{I}}^{(4)}$ where $\hat{\textrm{I}}^{(4)}$ is the $4\times4$ identity matrix.

\section{Spin-orbit interaction for semiconductor crystals\label{sec3}}

We now turn to semiconductor crystals and first derive the spatial operator $\hat{\bf\Lambda}$ defined in Eq.~(\ref{4}).

\noindent \textbf{(a) Periodic potential}

The first problem is to handle the periodicity of the $\mathcal{V}(\vr)$ potential felt by electrons in a  crystal, $\mathcal{V}(\vr)=\mathcal{V}(\vr+\vR_\ell)$ for any lattice vector $\vR_\ell$. The way to do it is to expand $\mathcal{V}(\vr)$ on reciprocal lattice vectors $\vQ$ that fulfill $e^{i\vQ\cdot\vR_\ell}=1$, namely
\be\label{12}
\mathcal{V}(\vr)=\sum_\vQ\mathcal{V}_\vQ e^{i\vQ\cdot\vr}
\ee
This gives $\nabla \mathcal{V}(\vr)=i \sum_\vQ  \vQ \,\,\mathcal{V}_\vQ \,e^{i\vQ\cdot\vr}$, which is obviously not a vector along $\vr$. So, the orbital angular momentum $\hat{\vL}=\vr\times\hat{\vp}$ is not going to appear in the spin-orbit interaction of semiconductor electrons.

\noindent \textbf{(b) Spatial basis for cubic crystals}

The next problem is to perform calculations using a spatial basis relevant to semiconductor electrons. From the Bloch theorem, we know that electrons in a periodic crystal are characterized by a band index $n$ and a wave vector $\vk$ that is quantized as $2\pi/L$ for a sample volume $L^3$, in order to fulfill the Born-von Karman boundary condition, $f(\textbf{r})=f(\textbf{r}+\textbf{L})$, which allows extending the crystal periodicity to a finite volume.

Valence electrons in the threefold level of a \textit{cubic} crystal have an additional spatial index that can be taken as $\mu=(x,y,z)$ along the crystal axes, in contrast to the  $(X,Y,Z)$ axes for atoms that can be chosen at will due to spherical symmetry. The valence-electron wave function then reads\cite{Merminbook}
\be\label{13}
\lan\vr|v,\mu,\vk\ran=\frac{e^{i\vk\cdot\vr}}{L^{3/2}}\,u_{v,\mu,\vk}(\vr)
\ee
 the Bloch function having the lattice periodicity $u_{v,\mu,\vk}(\vr)=u_{v,\mu,\vk}(\vr+\vR_\ell)$.

We want to find how the spin-orbit interaction splits the degeneracy of the threefold spatial level $(v,\mu)$, for $\vk=\bf0$, that is, at the valence band maximum\cite{9}. To handle the periodicity of the Bloch function, we also expand it on reciprocal lattice vectors: 
\be
\label{14}
\lan\vr|v,\mu,\vk=\v0\ran=\frac{1}{L^{3/2}}\,u_{v,\mu,\v0}(\vr)=\frac{1}{L^{3/2}}\sum_\vQ u_{v,\mu;\vQ}e^{i\vQ\cdot\vr}
\ee

\noindent \textbf{(c) $\hat{\bf\Lambda}^{(cb)}$ matrix in the $\mu$ basis}

$\bullet$ To calculate the $\hat{\bf\Lambda}^{(cb)}$ matrix elements in the $\mu$ basis, we first note, from Eq.~(\ref{14}), that 
\be
\lan \vr|\hat{\vp}|v,\mu,{\bf0}\ran=\frac{\hbar }{i}\nabla\lan \vr|v,\mu,{\bf0}\ran= \frac{\hbar}{L^{3/2}}\sum_\vQ  \vQ \,\,u_{v,\mu;\vQ}\,e^{i\vQ\cdot\vr}\label{15}
\ee
which leads, for the  $\mathcal{V}(\vr)$ potential given in Eq.~(\ref{12}) and  $\hat{\bf\Lambda}^{(cb)}$ defined in Eq.~(\ref{4}), to  
\bea\label{16}
&&\lan \vr| \hat{\bf\Lambda}^{(cb)}|v,\mu,{\bf0}\ran = 
\\
&&\,\,\,\,\,\,\,\,\,
\frac{i\hbar }{L^{3/2} }\sum_{\vQ_1}\sum_\vQ   \mathcal{V}_{\vQ_1}\, u_{v,\mu;\vQ}\,e^{i(\vQ+\vQ_1)\cdot\vr}\,\Big(\vQ_1\times \vQ\Big)
\nn
\eea
The matrix elements of the  $\hat{\bf\Lambda}^{(cb)}$ operator in the degenerate subspace $|v,\mu,{\bf0}\ran$ then read 
\bea
\label{17}
\lan v,\mu',{\bf0}|\hat{\bf\Lambda}^{(cb)} |v,\mu,{\bf0}\ran \!\!\!&=&\!\!\! i\hbar\sum_{\vQ_1}\sum_{\vQ'\vQ}   \mathcal{V}_{\vQ_1}\, u^*_{v,\mu';\vQ'} u_{v,\mu;\vQ}
\\
&&  \Big(\vQ_1\times \vQ\Big)    \int  \frac{d^3r }{L^3 } e^{i(\vQ+\vQ_1-\vQ')\cdot\vr}
\nn
\eea
The integral over $\vr$ imposes $\vQ'=\vQ+\vQ_1$; so, the above matrix element reduces to
\be
 i\hbar\sum_{\vQ'\vQ}   \mathcal{V}_{\vQ'-\vQ} \,u^*_{v,\mu';\vQ'} u_{v,\mu;\vQ}\,\Big(\vQ'\times\vQ\Big)\label{18}
\ee

$\bullet$ To go further, we note that for crystals with inversion symmetry, all valence-electron  wave functions are even, and all conduction-electron wave functions are odd\cite{Cardona}. So, the upper threefold valence level has an even parity, in spite of the fact that it has been misleadingly called P according to the atomic notation for a threefold level. This parity then imposes in Eq.~(\ref{14})
\be\label{19}
\lan \vr |v,\mu,{\bf0}\ran=\lan -\vr |v,\mu,{\bf0}\ran\Longleftrightarrow u_{v,\mu;\vQ}= u_{v,\mu;-\vQ}
\ee
which for a threefold level with cubic symmetry, leads us to take, due to cyclic permutations,
\be
u_{v,\mu;\vQ}=\frac{Q_xQ_yQ_z}{Q_\mu} G_{v,Q}\label{20}
\ee
where $G_{v,Q}$ depends on $Q=|\vQ|$ only.

$\bullet$ The last step is to show that all $\hat{\bf\Lambda}^{(cb)}$ matrix elements in the $|v,\mu,{\bf0}\ran$ basis are equal to zero, except
\be
\lan v,y,{\bf0}| \hat{\Lambda}_z^{(cb)} |v,x,{\bf0}\ran\equiv i\hbar \lambda_{cb}\label{21}
\ee
and its cyclic permutations, the constant $\lambda_{cb}$ being given by
\bea\label{22}
\lambda_{cb} &=& \sum_{\vQ'\vQ}   \mathcal{V}_{\vQ'-\vQ}\, G^*_{v,Q'}G_{v,Q}\nn\\
&&\times\big(Q'_xQ'_yQ'_z\big)\big(Q_xQ_yQ_z\big)\frac{Q'_xQ_y-Q'_yQ_x}{Q'_y Q_x}
\eea
which does not depend on $(x,y)$ due to cyclic permutations. The derivation of this key result, which requires 
\be\label{25}
\mathcal{V}(x,y,z)=\mathcal{V}(y,x,z)=\mathcal{V}(-x,y,z)
\ee 
as fulfilled by  the potential $\mathcal{V}(\vr)$ of a cubic crystal, is given in  Appendix \ref{app:A}.

\noindent \textbf{(d) Cubic spatial momentum} $\hat{\mathbfcal{L}}^{(cb)}$

By using Eq.~(\ref{21}), we can derive the components of the ``spatial momentum'' 
\be\label{23}
\hat{\mathbfcal{L}}^{(cb)}=\frac{\hat{\bf\Lambda}^{(cb)}}{\lambda_{cb}}
\ee 
in the $|v,\mu,{\bf0}\ran$ basis. From them, we find that
the matrix representations of the $\hat{\mathbfcal{L}}^{(cb)}$ components read as 
\bea 
\hat{\mathcal{L}}_x^{(cb)}=\hbar \begin{pmatrix}
0 & 0&0 \\
0 & 0&-i \\
0 & i&0
\end{pmatrix}_\mu\qquad \hat{\mathcal{L}}_y^{(cb)}=\hbar \begin{pmatrix}
0 & 0&i \\
0 & 0&0 \\
-i & 0&0
\end{pmatrix}_\mu \nn\\
 \hat{\mathcal{L}}_z^{(cb)}=\hbar \begin{pmatrix}
0 & -i&0 \\
i & 0&0 \\
0 & 0&0
\end{pmatrix}_\mu\hspace{2cm}\label{24}
\eea
We can check that like for the orbital angular momentum $\hat{\vL}$, these components fulfill $\left[\hat{\mathcal{L}}_x^{(cb)},\hat{\mathcal{L}}_y^{(cb)}\right]_-=i\hbar \hat{\mathcal{L}}_z^{(cb)}$, and $(\hat{\mathbfcal{L}}^{(cb)})^2=(1\cdot2)\hbar^2\,\hat{\textrm{I}}^{(3)}$.

$\bullet$ We now anticipate that, when the spin is included, the appropriate spatial basis will not be $|v,\mu,{\bf0}\ran$, but $|v,\mathcal{L}_z,{\bf0}\ran$ with $\mathcal{L}_z=(\pm1,0)$, that we define according to the Landau-Lifshitz phase factor for spherical harmonics\cite{Landau01} as
\bea
|v,\pm1_z,{\bf0}\ran &=&\frac{\mp i |v,x,{\bf0}\ran+ |v,y,{\bf0}\ran}{\sqrt{2}}\label{26}\\
|v,0_z,{\bf0}\ran &=& i|v,z,{\bf0}\ran \label{27}
\eea
By using Eq.~(\ref{24}), we find that the matrix elements of the $\hat{\mathbfcal{L}}^{(cb)}$ operator in this basis are given by
\bea
\lan v,\eta'_z,{\bf0}| \hat{\mathcal{L}}_{z}^{(cb)} |v,\eta_z,{\bf0}\ran &=& \hbar  \frac{\eta'_z+\eta_z}{2}\label{28}\\
\lan v,\eta_z,{\bf0}| \hat{\mathcal{L}}_{\pm}^{(cb)} |v,0_z,{\bf0}\ran &=& \hbar  \frac{1\pm\eta_z}{\sqrt{2}}\label{29}
\eea
for $\eta=\pm1$. Using them, we get the matrix representations of these operators in the $|v,\mathcal{L}_z,{\bf0}\ran$ basis as  
\bea\label{30}
\hat{\mathcal{L}}_z^{(cb)}=\hbar\begin{pmatrix}
1 & 0&0 \\
0 & 0&0 \\
0 & 0&-1
\end{pmatrix}_{1;z}
 \\
\quad
 \hat{\mathcal{L}}_+^{(cb)}=(\hat{\mathcal{L}}_-^{(cb)})^\dag=\hbar \sqrt{2}\, \begin{pmatrix}
0 & 1&0 \\
0 & 0&1 \\
0 & 0&0
\end{pmatrix}_{1;z}
 \nn
\eea
These matrices have the same form as the ones, in Eq.~(\ref{6}), for the orbital angular momentum $\hat{\vL}$, provided that the atomic axes $(X,Y,Z)$ are chosen along the cubic axes $(x,y,z)$.

\noindent \textbf{(e) Cubic hybrid momentum} $\hat{\mathbfcal{J}}^{(cb)}$

 $\bullet$ The last step is to introduce the spin. The spin-orbit interaction given in Eq.~(\ref{2}) reads
\be\label{31}
\hat{H}_{so}=\lambda_{so}\hat{\bf\Lambda}^{(cb)}\cdot \hat{\vS}= \lambda_{so}\,\lambda_{cb}\,\,\hat{\mathbfcal{L}}^{(cb)}\cdot \hat{\vS}
\ee 
 that we rewrite as 
 \be\label{32}
 \hat{H}_{so}=\lambda_{so}\lambda_{cb}\left(\frac{\hat{\mathcal{L}}_+^{(cb)} \hat{S}_-+\hat{\mathcal{L}}_- ^{(cb)}\hat{S}_+}{2}+\hat{\mathcal{L}}_z^{(cb)} \hat{S}_z\right)
 \ee
 with $\hat{S}_\pm=\hat{S}_x\pm i \hat{S}_y$.
 
 Using Eqs.~(\ref{28},\ref{29}), we can deduce the matrix elements of the $\hat{\mathbfcal{L}}^{(cb)}\cdot\hat{\vS}$ operator in the six-state basis $|(1_z,0_z,{-}1_z)\ran\otimes |1/2_z\ran$ and $|(1_z,0_z,{-}1_z)\ran\otimes |{-}1/2_z\ran$. The corresponding matrix appears as
 \be\label{33}
\hat{\mathbfcal{L}}^{(cb)}\cdot\hat{\vS}=\frac{\hbar^2}{2}\begin{pmatrix}
1 & 0 &0 & 0 &0 &0 \\
0 & 0 &0 & \sqrt{2} &0 &0 \\
0 & 0 &-1 & 0 & \sqrt{2} &0 \\
0 & \sqrt{2} &0 & -1 & 0 &0 \\
0 & 0 &\sqrt{2} & 0 & 0 &0 \\
0 & 0 &0 & 0 & 0 &1 \\
\end{pmatrix}
 \ee
 It can be made block-diagonal by interchanging the  states $|{-}1_z\ran\otimes |1/2_z\ran$ and $|1_z\ran\otimes |{-}1/2_z\ran$, to get
  \be\label{34}
\hat{\mathbfcal{L}}^{(cb)}\cdot\hat{\vS}=\frac{\hbar^2}{2}\begin{pmatrix}
1 & 0 &0   & 0 &0 &0 \\
0 & 0 &\sqrt{2}  & 0 &0 &0 \\
0 &\sqrt{2} & -1 &  0  & 0 &0 \\
0 & 0 &0  & -1  & \sqrt{2} &0 \\
0 & 0 &0  &\sqrt{2} & 0 &0 \\
0 & 0 &0  & 0 & 0 &1 \\
\end{pmatrix}
 \ee
 
 The eigenvalues of the upper-left $(3\times3)$ submatrix follow from the determinant
\be\label{35}
0=\left|\begin{matrix}
1{-}u & 0 &0  \\
0 & {-}u &\sqrt{2} \\
0 &\sqrt{2} & {-}1{-}u
\end{matrix}\right|=(1-u)(u^2+u-2)
\ee
And similarly for the lower-right $(3\times3)$ submatrix.
Their eigenvalues are $u=(1,1,-2)$.

 The basis is made of the states $|\eta_z\ran\otimes |\eta_z/2\ran$, $ |0_z\ran\otimes |\eta_z/2\ran$ and $|\eta_z\ran\otimes |-\eta_z/2\ran$, with $\eta=+1$ for the upper-left $(3\times3)$ submatrix, and $\eta=-1$ for the lower-right $(3\times3)$ submatrix. The four eigenstates that correspond to the $u=1$ eigenvalue, that is, a spin-orbit shift equal to $(\hbar^2/2)\lambda_{so}\lambda_{cb}$, read
 \bea
 |\pm1_z\ran\otimes
|(\pm\frac{1}{2})_z\ran 
 &\equiv& |\pm A\ran\label{36}\\
  \frac{|\pm1_z\ran\otimes\left|(\mp\frac{1}{2})_z\right\ran +\sqrt{2}|0_z\ran\otimes\left|(\pm\frac{1}{2})_z\right\ran }{\sqrt{3}}&\equiv& |\pm A'\ran\label{37}
 \eea
 while the two   eigenstates that correspond to the $u=-2$ eigenvalue, that is, a spin-orbit shift equal to $-\hbar^2\lambda_{so}\lambda_{cb}$, read 
 \be\label{38}
 \pm\frac{\sqrt{2}|\pm1_z\ran\otimes\left|(\mp\frac{1}{2})_z\right\ran -|0_z\ran\otimes\left|(\pm\frac{1}{2})_z\right\ran }{\sqrt{3}}\equiv |\pm B\ran
 \ee
We see that the $|\pm A\ran$ and $|\pm A'\ran$ eigenstates are just the atomic states $|j=3/2,j_Z=\pm3/2\ran$ and $|j=3/2,j_Z=\pm1/2\ran$, while $|\pm B\ran$ correspond to $|j=1/2,j_Z=\pm1/2\ran$,
for $Z$ chosen along $z$.

 $\bullet$ This remark leads us to introduce the vector operator $\hat{\mathbfcal{J}}^{(cb)}$, formally defined as
 \be\label{39}
 \hat{\mathbfcal{J}}^{(cb)}=\hat{\mathbfcal{L}}^{(cb)}+\hat{\vS}
 \ee
 where the operators $\hat{\mathbfcal{L}}^{(cb)}$ and $\hat{\vS}$ respectively act in the real space and the spin space. From $\left[\hat{\mathcal{L}}_x^{(cb)},\hat{\mathcal{L}}_y^{(cb)}\right]_-=i\hbar \hat{\mathcal{L}}_z^{(cb)}$ and $\left[\hat{S}_x,\hat{S}_y\right]_-=i\hbar \hat{S}_z$, we readily get
 \be\label{40}
 \left[\hat{\mathcal{J}}_x^{(cb)},\hat{\mathcal{J}}_y^{(cb)}\right]_-=i\hbar \hat{\mathcal{J}}_z^{(cb)}
 \ee
 with cyclic permutations.
 
As $(\hat{\mathbfcal{J}}^{(cb)})^2=(\hat{\mathbfcal{L}}^{(cb)})^2+\hat{\vS}^2+2\hat{\mathbfcal{L}}^{(cb)}\cdot\hat{\vS}$, while $(\hat{\mathbfcal{L}}^{(cb)})^2$ gives $(1\cdot2)\hbar^2\,\hat{\textrm{I}}^{(3)}$ on any spatial states $|\mathcal{L}_z\ran$ with $\mathcal{L}_z=(\pm1,0)$, we find that the operator $(\hat{\mathbfcal{J}}^{(cb)})^2$ is diagonal in the $(|\pm A\ran,|\pm A'\ran)$ subspace that corresponds to the  $(\hbar^2/2)\lambda_{so}\lambda_{cb}$ spin-orbit shift; the associated eigenvalue, equal to 
 \be\label{41}
1\cdot 2 \hbar^2+\frac{1}{2}\frac{3}{2}\hbar^2+2\,\frac{\hbar^2}{2}=\frac{3}{2}\frac{5}{2}\hbar^2
 \ee
  just corresponds to $\mathcal{J}(\mathcal{J}+1)\hbar^2$ for $\mathcal{J}=3/2$. In the same way, we find that the $(\hat{\mathbfcal{J}}^{(cb)})^2$ eigenvalue in the $|\pm B\ran$ subspace that corresponds to the  $(-\hbar^2)\lambda_{so}\lambda_{cb}$ spin-orbit shift, is equal to  
\be\label{42}
1\cdot 2 \hbar^2+\frac{1}{2}\frac{3}{2} \hbar^2 -2\,\hbar^2=\frac{1}{2}\frac{3}{2}\hbar^2
 \ee
 which just corresponds to $\mathcal{J}(\mathcal{J}+1)\hbar^2$ for $\mathcal{J}=1/2$.
 
 Moreover, we note that the $|\pm A\ran$, $|\pm A'\ran$, and $|\pm B\ran$ states also are  eigenstates of  $\hat{\mathcal{J}}_z^{(cb)}=\hat{\mathcal{L}}_z^{(cb)}+\hat{S}_z$,
 \begin{subeqnarray}\label{43}
 \hat{\mathcal{J}}_z^{(cb)}|\pm A\ran&=&\pm\frac{3}{2}\hbar|\pm A\ran\\
   \hat{\mathcal{J}}_z^{(cb)}|\pm A'\ran&=&\pm\frac{1}{2}\hbar|\pm A'\ran\\
   \hat{\mathcal{J}}_z^{(cb)}|\pm B\ran&=&\pm\frac{1}{2}\hbar|\pm B\ran
 \end{subeqnarray}
 The fact that these states are eigenstates of the $((\hat{\mathbfcal{J}}^{(cb)})^2,\hat{\mathcal{J}}_z^{(cb)})$ operators leads to the following identification
 \begin{subeqnarray}
|\pm A\ran&\equiv&\left|\mathcal{J}=\frac{3}{2},\mathcal{J}_z=\pm\frac{3}{2}\right\ran\slabel{44}\\
 |\pm A'\ran&\equiv&\left|\mathcal{J}=\frac{3}{2},\mathcal{J}_z=\pm\frac{1}{2}\right\ran\slabel{45}\\
 |\pm B\ran&\equiv&\left|\mathcal{J}=\frac{1}{2},\mathcal{J}_z=\pm\frac{1}{2}\right\ran\slabel{46}
\end{subeqnarray}

 From them, it is easy to obtain the matrix representation of the vector operator $\hat{\mathbfcal{J}}^{(cb)}$ in the $(|\pm A\ran,|\pm A'\ran)$ subspace and  in the  $|\pm B\ran$ subspace. In particular, the $ \hat{\mathcal{J}}_z^{(cb)}$ components read in these subspaces as
 \be
 \hat{\mathcal{J}}_z^{(cb)}=\hbar\begin{pmatrix}
\frac{3}{2} & 0&0 &0 \\
0 & \frac{1}{2}&0 &0\\
0 & 0&-\frac{1}{2}&0\\
0 & 0&0&-\frac{3}{2}
\end{pmatrix}_{\frac{3}{2};z}\quad  \hat{\mathcal{J}}_z^{(cb)}=\hbar\begin{pmatrix}
\frac{1}{2} &0\\
 0&-\frac{1}{2}
\end{pmatrix}_{\frac{1}{2};z}
 \ee
 
\section{Phase factor between valence-electron and hole operators\label{sec4}}

Relativistic quantum theory\cite{Berestetskii} gives the link between the destruction operator of a particle with quantum indices $(j,j_Z)$ and the creation  operator of its antiparticle\cite{Landau01}, as
\be
\label{47}
\hat{a}_{j,j_Z}=(-1)^{j-j_Z} \hat{b}^\dag_{j,-j_Z}\,
\ee

To provide a strong physical support to these cubic operators $\hat{\mathbfcal{L}}^{(cb)}$ and $\hat{\mathbfcal{J}}^{(cb)}$, let us show that, although the hole is definitely not a na\"{i}ve antiparticle of the valence electron due to its interactions with the electrons that remain in the valence band\cite{Monicbook}, so that the hole is a many-body object in itself, the above relation is still valid for cubic semiconductors\cite{SS2021}, provided that we label the valence electron states along the eigenvalues of the cubic spatial momentum $\hat{\mathbfcal{L}}^{(cb)}$ or the cubic hybrid momentum $\hat{\mathbfcal{J}}^{(cb)}$ defined in the preceding section.

$\bullet$ The derivation of this result relies on
\be
\label{48}
\hat{a}_{(\pm \frac{1}{2})_z}=\pm \,\hat{b}^\dag_{(\mp\frac{1}{2})_z}
\ee
that follows from Eq.~(\ref{47}) taken for the spin, that is, $j=1/2$, and 
\be\label{49}
\hat{a}_\mu=\hat{b}^\dag_\mu \,\,\,\,\, \,\,\,\,\,\,\textrm{for} \,\,\,\,\,\,\,\,\,      \mu=(x,y,z)
\ee
 as imposed by cyclic permutations in a cubic crystal.

$\bullet$ Equations (\ref{26},\ref{27}) then give
\bea
\hat{a}_{\pm1_z}&=& \frac{\pm i \hat{a}_x +\hat{a}_y}{\sqrt{2}}=\frac{\pm i \hat{b}^\dag_x+\hat{b}^\dag_y}{\sqrt{2}}=\hat{b}^\dag_{\mp1_z}\,\label{ap_eth_21a}\\
\hat{a}_{0_z}&=& -i \hat{a}_z=-i\hat{b}^\dag_z=-\hat{b}^\dag_{0_z}\,\label{ap_eth_21b}
\eea
which agree with Eq.~(\ref{47}) for $j=1$ and $j_z=(\pm1,0)$ with $Z$ taken along $z$.

$\bullet$ When turning to the four eigenstates defined in Eqs.~(\ref{36},\ref{37}), the above equations give for $|\pm A\ran$ seen as $(\mathcal{J}=3/2,\mathcal{J}_z=\pm3/2)$,
\be
\hat{a}^\dag_{\frac{3}{2},(\pm \frac{3}{2})_z} =\hat{a}^\dag_{\pm 1_z,( \pm\frac 1 2)_z}=\pm\hat{b}_{\mp 1_z, (\mp\frac 1 2)_z}=\pm \hat{b}_{\frac{3}{2},(\mp \frac{3}{2})_z}\label{app:etoh:a+323eta2=}
\ee
while for $|\pm A'\ran$ seen as $(\mathcal{J}=3/2,\mathcal{J}_z=\pm1/2)$, we get 
\bea
\hat{a}^\dag_{\frac{3}{2},(\pm\frac{1}{2})_z}&=&\frac{\hat{a}^\dag_{\pm1_z, (\mp \frac{1}{2})_z}+ \sqrt{2}\,\hat{a}^\dag_{0_z, (\pm \frac 1 2)_z}  }{\sqrt{3}}\nn\\
&=& \frac{(\mp)\hat{b}^\dag_{\mp1_z, (\pm \frac{1}{2})_z}-(\pm) \sqrt{2}\,\hat{b}^\dag_{0_z, (\mp \frac 1 2)_z}  }{\sqrt{3}}\nn\\
&=& \mp \hat{b}^\dag_{\frac{3}{2},( \mp \frac 1 2)_z}
\eea
We again see that the change  from valence-electron destruction operator to hole creation operator agrees with Eq.~(\ref{47}).

$\bullet$ We can check that this is also true for the $|\pm B\ran$ states of Eq.~(\ref{38}), seen as $(\mathcal{J}=1/2,\mathcal{J}_z=\pm1/2)$. Indeed,
\bea
\hat{a}^\dag_{\frac{1}{2}, (\pm \frac{1}{2})_z}\!\!\!&=&\!\!\!\pm\frac{\sqrt{2}~ \hat{a}^\dag_{\pm 1_z, (\mp\frac{1}{2})_z}-\hat{a}^\dag_{0_z,(\pm\frac 1 2)_z}}{\sqrt{3}}
\nn\\
&=&\pm \frac{(\mp)\sqrt{2}~ \hat{b}^\dag_{\mp1_z,( \pm\frac{1}{2})_z}+(\pm) \hat{b}^\dag_{0_z,(\mp\frac 1 2)_z}}{\sqrt{3}}\nn\\
&=&\pm  \hat{b}^\dag_{\frac{1}{2},(\mp\frac{1}{2})_z}
\eea

\section{Possible extension}\label{sec5}

We wish to stress that the definition of the vector operator $\hat{\mathbfcal{L}}$ given in Eq.~(\ref{4}) is completely general: it only reads in terms of the electron momentum operator $\hat{\vp}$ and the internal electrostatic potential $\mathcal{V}(\vr)$ felt by semiconductor electrons, without any restriction on the potential symmetry. When this potential has the cubic symmetry, it does not come as a surprise to find that $\hat{\mathbfcal{L}}^{(cb)}$ has exactly the same matrix representation as the orbital angular momentum  $\hat{\vL}$ for the $\ell=1$ atomic level, because the $(x,y,z)$ orthogonal axes play the same role for a sphere and a cube. 

We can go further and ask whether the $\hat{\mathbfcal{L}}$ concept can be extended to a crystal having less symmetry than the cubic symmetry, for example when the crystal axes $(x,y,z)$ are not equivalent, or even not orthogonal. The identity between the $\hat{\mathbfcal{L}}$ and $\hat{\vL}$ matrices will not exist anymore, but the $\hat{\mathbfcal{J}}$ eigenstates, for $\hat{\mathbfcal{J}}$ constructed in the same way out of $\hat{\mathbfcal{L}}$, may still provide a convenient way to derive the semiconductor eigenstates in the presence of spin-orbit interaction, without having to resort to the group theory which is physically obscure when turning to the double group, as required to handle the spin-orbit interaction. The possible extension of the present work to periodic systems with other potential symmetries deserves further investigation; it just constitutes the starting point to move forward.

\section{Conclusion}

 Through a  microscopic procedure, we here show that the spin-orbit interaction in semiconductors has the same $\hat{\mathbfcal{L}}\cdot\hat{\vS}$ form as for atomic electrons, with $\hat{\mathbfcal{L}}$ being the analog of the orbital angular momentum $\hat{\vL}$, as supported by their identical matrix representations in the atomic basis and the Bloch-state basis for cubic semiconductors. Our work thus provides the long-missed support for using atomic notations to label valence electron states in GaAs-like semiconductors, in spite of the fact that the electrostatic potential felt by semiconductor electrons is not spherical but periodic. Up to now, the only clean classification for these electrons relied on the group theory, which is definitively correct but overly heavy when dealing with a crystal symmetry as simple as cubic.

  As a strong support to this labeling, we also show that the transformation from valence electron to hole operators  in a cubic semiconductor, has the same phase factor as the one from particle to antiparticle in relativistic quantum theory. This provides a secure way to describe semiconductor physics in terms of electrons and holes through effective electron-hole Hamiltonian\cite{KL,Monicbook} and effective coupling to the electromagnetic field\cite{MC1990,Hartmutbook}. 
  
 \section{Acknowledgment}   
It is our pleasure to acknowledge constructive discussions with the Referee and with Benoit Eble, about the appropriate way to name $\hat{\mathbfcal{L}}$ and $\hat{\mathbfcal{J}}$.

\appendix

\section{Derivation of Eq.~(\ref{21})\label{app:A}} 

We here derive Eq.~(\ref{21}) starting from Eq.~(\ref{18}), namely
 \bea
&&\lan v,\mu',{\bf0}|\hat{\bf\Lambda}^{(cb)} |v,\mu,{\bf0}\ran
   \\
&&\,\,\,\,\,\,\,\,\,\,\,\, = i\hbar\sum_{\vQ'\vQ}   \mathcal{V}_{\vQ'-\vQ} \,u^*_{v,\mu';\vQ'} u_{v,\mu;\vQ}\,\Big(\vQ'\times\vQ\Big)\label{A1}
 \nn
\eea

 {\bf(i)} To do it, we first consider $\mu'=\mu$. By using Eq.~(\ref{20}), the above equation gives the $\hat{\Lambda}_x^{(cb)}$ component as
\bea
\lan v,\mu,{\bf0}|\hat{\Lambda}_x^{(cb)} |v,\mu,{\bf0}\ran = i\hbar\sum_{\vQ'\vQ}   \mathcal{V}_{\vQ'-\vQ}\, G^*_{v,Q'}G_{v,Q}\label{A2}
\hspace{1cm}\\
\times\left(\frac{Q'_xQ'_yQ'_z}{Q'_\mu} \frac{Q_xQ_yQ_z}{Q_\mu}\right)
\,\big(Q_y' Q_z-Q_yQ_z'\big)\nn
\eea
When reversing the direction of the \textbf{z} axis, $(Q_z,Q'_z)$ change into $(-Q_z,-Q'_z)$; so, the first factor $\left(\frac{Q'_xQ'_yQ'_z}{Q'_\mu} \frac{Q_xQ_yQ_z}{Q_\mu}\right)$ keeps its sign whatever $\mu$, while the second factor $(Q_y' Q_z-Q_yQ_z')$ changes sign. Since  a potential with cubic symmetry is not affected by reversing the direction of the \textbf{z} axis, namely,  $\mathcal{V}(x,y,z)=\mathcal{V}(x,y,-z)$, we end with $\lan v,\mu,{\bf0}|\hat{\Lambda}_x ^{(cb)}|v,\mu,{\bf0}\ran=\,-\,\lan v,\mu,{\bf0}|\hat{\Lambda}_x^{(cb)} |v,\mu,{\bf0}\ran$, which proves that this matrix element is equal to zero.

The same is true for $\hat{\Lambda}_y^{(cb)}$ by reversing the direction of the \textbf{x} 
axis, and for $\hat{\Lambda}_z^{(cb)}$ by reversing the direction of the \textbf{y} 
axis. So, we end with
\bea
\lan v,\mu,{\bf0}|\hat{\Lambda}_{\mu''}^{(cb)} |v,\mu,{\bf0}\ran=0
\eea
whatever $\mu$ and $\mu''$.

 {\bf(ii)} We now consider $\mu'\neq\mu$. By reversing the directions of the $\mu'$ axis and the $\mu$ axis, the cubic potential still stays unchanged and the factor $\left(\frac{Q'_xQ'_yQ'_z}{Q'_{\mu'}} \frac{Q_xQ_yQ_z}{Q_\mu}\right)$ still keeps its sign. In order for the matrix elements of $\hat{\Lambda}_{\mu''}^{(cb)}$ to also keep its sign, $\mu''$ must be different from $(\mu', \mu)$, like $x$ for $(y,z)$. The component of $\big(\vQ'\times\vQ\big)$ along $\mu''$  in Eq.~(\ref{A1}) then reads in terms of $Q'_{\mu'} Q_\mu$. So,  to get nonzero matrix elements, that is, 
 \bea
\lan v,\mu',{\bf0}|\hat{\Lambda}_{\mu''}^{(cb)} |v,\mu,{\bf0}\ran\neq0
\eea
we must have the $(\mu,\mu',\mu'')$ indices all different, the values of the above matrix elements being independent of these indices due to cubic symmetry.

\end{document}